\documentclass[journal=jacsat,manuscript=article]{achemso}

\usepackage[version=3]{mhchem} 
\usepackage{amsmath}
\usepackage{braket}
\usepackage{xcolor}
\usepackage{amsthm,amsmath,amssymb}
\usepackage{physics}

\usepackage{accents}
\newlength{\dhatheight}
\newcommand{\doublehat}[1]{%
    \settoheight{\dhatheight}{\ensuremath{\hat{#1}}}%
    \addtolength{\dhatheight}{-0.35ex}%
    \hat{\vphantom{\rule{1pt}{\dhatheight}}%
    \smash{\hat{#1}}}}

\author{Yu Wang}
\affiliation{Department of Chemistry, School of Science, Westlake University, Hangzhou 310024 Zhejiang, China}
\altaffiliation
{Institute of Natural Sciences, Westlake Institute for Advanced Study, Hangzhou 310024 Zhejiang, China}

\author{Vahid Mosallanejad}
\affiliation{Department of Chemistry, School of Science, Westlake University, Hangzhou 310024 Zhejiang, China}
\altaffiliation
{Institute of Natural Sciences, Westlake Institute for Advanced Study, Hangzhou 310024 Zhejiang, China}

\author{Wei Liu}
\affiliation{Department of Chemistry, School of Science, Westlake University, Hangzhou 310024 Zhejiang, China}
\altaffiliation
{Institute of Natural Sciences, Westlake Institute for Advanced Study, Hangzhou 310024 Zhejiang, China}

\author{Wenjie Dou}
\email{douwenjie@westlake.edu.cn}
\affiliation{Department of Chemistry, School of Science, Westlake University, Hangzhou 310024 Zhejiang, China}
\altaffiliation
{Institute of Natural Sciences, Westlake Institute for Advanced Study, Hangzhou 310024 Zhejiang, China}

\title
  {Nonadiabatic dynamics near metal surfaces with periodic drivings: A generalized surface hopping in Floquet representation}

\begin{document}








\begin{abstract}
    With light-matter interaction extending into strong regime, as well as rapid development of laser technology, systems subjecting to a time-periodic perturbation are attracted broad attention.  Floquet theorem and Floquet time-independent Hamiltonian are powerful theoretical framework to investigate the systems subjecting to time-periodic drivings. In this study, we extend the previous generalized SH algorithm near metal surface (J. Chem. Theory Comput. 2017, 13, 6, 2430-2439) to the Floquet space, and hence, we develop a generalized Floquet representation based surface hopping (FR-SH) algorithm.  Here, we consider open quantum system with fast drivings. We expect that the present algorithm will be useful for understanding the chemical processes of molecules under time-periodic drivings near the metal surface. 
\end{abstract}

\section{1. Introduction}
The Born-Oppenheimer (BO) approximation is established based on the ratio of the mass of a nucleus to the mass of an electron so that nuclear motion is decoupled from electronic dynamics. However, when it comes to electronic excitation transfer or any form of electronic relaxation, BO dynamics break down such that one must take into account the coupling between the nuclear motion and electronic transitions\cite{subotnik2016understanding,agostini2019different}.
When concerning molecule-metal interfaces, electrons from the metal are much easier to excite than for an isolated molecule, such that nonadiabatic dynamics are inevitable\cite{dou2020nonadiabatic}.
At molecule-metal interfaces, there are numerous chemical setups including heterogeneous catalysis,\cite{cui2018bridging,zhong2020state,bavykina2020metal} chemisorption,\cite{huber2019chemical,cai2021catalytic} and molecular junctions\cite{gu2021tuning,liu2021charge}.
For nonadiabatic dynamics, there are several numerical exact solutions, including numerical renormalization group (NRG) techniques,\cite{bulla2008numerical,costi1994transport} multi-configuration time dependent Hartree (MCTDH),\cite{thoss2007correlated} hierarchical quantum master equation (HQME),\cite{schinabeck2016hierarchical,xu2019non} and quantum Monte Carlo (QMC)\cite{muhlbacher2008real}.
However, these methods cannot deal with large degrees of freedom (DoFs) in a metal.
Several years ago, Dou et al. have developed a surface hopping (SH) algorithm near the metal surface which is based on a simple classical master equation (CME).\cite{dou2015surface,dou2015surface3}
Such SH algorithm treats all metallic electrons implicitly, and works in the limit of weak molecule-metal interactions.
For a realistic molecule with more than one orbital near metal surface, Dou et al. developed a generalized SH algorithm in which the quantum-classical Liouville equation (QCLE) was embedded inside a CME.\cite{dou2016many,dou2017generalized}
In this study, we extend the generalized SH algorithm\cite{dou2017generalized} to treat Floquet engineered molecules near the metal surface.

Over the last few decades, advancements of high-power and short-pulse laser technologies have greatly facilitated the exploration of novel atomic and molecular properties.\cite{manakov1986atoms,nerush2011laser,fulop2020laser} 
It is also demonstrated that even without external light, vacuum fluctuations in the cavity can strongly couple with the matter when electronic or vibrational transitions of the matter are resonant with the cavity mode\cite{garcia2021manipulating}. 
Such strong light-matter interaction has been achieved to modify the nature of the matter, which resulting in novel properties including electronic conductivity,\cite{orgiu2015conductivity,krainova2020polaron} energy transfer probability,\cite{coles2014polariton,zhong2016non} nonlinear optical response,\cite{daskalakis2014nonlinear,goblot2019nonlinear,zhao2022nonlinear} chemical reaction,\cite{hutchison2012modifying,herrera2016cavity} and so forth.
It is thus necessary to theoretically investigate how matter properties can be modified by these strong external field.
As systems are subject to a strong time-periodic external stimulus, the standard nonlinear perturbation theory\cite{mukamel1999principles} becomes inaccurate.
The Floquet theory is widely used in theoretical investigations, because it allows the reduction of the periodical or quasiperiodical time-dependent Schr\"{o}dinger equation into a set of time-independent coupled equations.\cite{kohler2018dispersive,ivanov2021floquet, engelhardt2021dynamical}

Floquet based nonadiabatic dynamics have been developed for both closed and open systems. For example, Schir{\`o} et al. developed a Floquet coupled-trajectory mixed quantum–classical (F-CT-MQC) algorithm for excited-state molecular dynamics simulations of systems subject to an external periodic drive. \cite{schiro2021quantum}
Sato et al. employed the Maxwell–Bloch equation to investigate basic properties of nonequilibrium steady states of periodically-driven open quantum systems.
\cite{sato2020floquet}
Li et al. developed a Floquet engineering scattering formalism that relies on a systematic high-frequency expansion of the
scattering matrix.\cite{li2018floquet}
Fiedlschuster et al used a Floquet based SH algorithm to describe the dynamics of one positively charged hydrogen in strong field.\cite{fiedlschuster2016floquet}
Nafari et al. employed Floquet master equation to study energy transfer from quantum system to two thermal baths.
\cite{qaleh2022enhancing}
Chen et al. investigated different approaches to derive the proper Floquet-based quantum–classical Liouville equation (F-QCLE) for laser-driven electron-nuclear dynamics.\cite{chen2020proper}
Wang et al. developed a Floquet surface hopping and Floquet electronic friction for one level near metal surface\cite{wang2023nonadiabatic,wang2023EF}.

In this study, we develop a generalized Floquet representation based SH (FR-SH) algorithm near the metal surface in an extended Hilbert space.
We organize the paper as follows: in section 2, we introduce the FR-QCLE-CME and FR-SH algorithm. In section 3, we discuss the results of FR-SH and FR-QME. We conclude in section 4.

\section{2. Theory}
\subsection{A. Equation of motion in Floquet representation}
To be explicit, we divide the general total Hamiltonian into three parts: the system $H_s$ that is driven by external drivings, the bath $H_b$, and the system-bath coupling $H_c$
\begin{equation}
    \hat{H}(t) = \hat{H}_s(t) + \hat{H}_b + \hat{H}_c
\end{equation}
\begin{equation}\label{Hs}
    \hat{H}_s(t) = \sum_{ij}h_{ij}(\hat{\boldsymbol{R}},t)\hat{d}^+_i \hat{d}_j + U_0(\hat{\boldsymbol{R}}) + \sum_{\alpha}\frac{\hat{\boldsymbol{P}}_{\alpha}^2}{2m_{\alpha}}
\end{equation}
\begin{equation}
    \hat{H}_b = \sum_{k}\epsilon_k \hat{c}_k^+ \hat{c}_k
\end{equation}
\begin{equation}
    \hat{H}_c = \sum_{ki}V_{ik}(\hat{c}_k^+ \hat{d}_i + \hat{d}^+_i \hat{c}_k)
\end{equation}
Here $\hat{d}^+_i(\hat{d_i})$ is the creation (annihilation) operator for the $i$-th electronic orbital of the molecule; $\hat{c}^+_k(\hat{c}_k)$ is the creation (annihilation) operator for the $k$-th electronic orbital of the metal surface. $\hat{\boldsymbol{R}}$ and $\hat{\boldsymbol{P}}$ are nuclear positions and momenta, respectively ($\alpha$ represents the degree of freedom of nuclei).
$m_{\alpha}$ is the nuclear mass.
$U_0(\hat{\boldsymbol{R}})$ is the diabatic nuclear potential for the unoccupied state. 
We consider a system under the periodic driving so that $h_{ij}(\hat{\boldsymbol{R}},t+T)=h_{ij}(\hat{\boldsymbol{R}},t)$, where $T$ is the driving period. $V_{ik}$ is the coupling between the molecular orbital $\hat{d}_i$ and the metallic orbital $\hat{c}_k$. Under the wide band approximation, the hybridization function $\Gamma_{ij}(\epsilon)$ measures the strength of system-bath coupling which is independent of $\epsilon$, and is given by:
\begin{equation}
    \Gamma_{ij}(\epsilon) = 2\pi\sum_k V_{ik}V_{jk}\delta(\epsilon-\epsilon_k)=\Gamma_{ij}
\end{equation}

The equation of motion (EOM) for the total density operator follows Liouville-von Neumann (LvN) equation:
\begin{equation}\label{EOM0}
    \frac{\partial}{\partial t}\hat{\rho}(t) = -\frac{i}{\hbar}[\hat{H}(t),\hat{\rho}(t)]
\end{equation}
In the weak system-bath coupling limit, we can reduce the LvN equation to the Redfield equation
, in which the EOM of the reduced system density operator $\hat{\rho}_s$(t) is\cite{dou2016many,dou2017generalized},
\begin{equation}\label{EOM}
    \frac{\partial}{\partial t}\hat{\rho}_s(t) = -\frac{i}{\hbar}[\hat{H}_s,\hat{\rho}_s(t)] - \doublehat{\mathcal{L}}_{bs}\hat{\rho}_s(t)
\end{equation}
here, the superoperator $\doublehat{\mathcal{L}}_{bs}$ includes information of system-bath couplings, which is,
\begin{equation}
\begin{split}
    \doublehat{\mathcal{L}}_{bs}\hat{\rho}_s(t) = \frac{1}{\hbar^2}\int_0^{\infty}d\tau e^{-i\hat{H}_st/\hbar}Tr_b([\hat{H}_{Ic}(t), [\hat{H}_{Ic}(t-\tau),e^{i\hat{H}_st/\hbar}\hat{\rho}_s(t)e^{-i\hat{H}_st/\hbar}\otimes\hat{\rho}_b^{eq}]])e^{i\hat{H}_st/\hbar}
\end{split}
\end{equation}
where $Tr_b$ indicates tracing over the bath ($i.e.$ the metal surface) degree of freedoms (DoFs), and $\hat{\rho}_b^{eq}$ is the bath equilibrium density operators. Note that $\hat{H}_{Ic}(t)$ in the above equation is in the interaction picture, $\hat{H}_{Ic}(t)=e^{i(\hat{H}_s+\hat{H}_b)t/\hbar}\hat{H}_ce^{-i(\hat{H}_s+\hat{H}_b)t/\hbar}$. We refer to Eq. (\ref{EOM}) as the quantum master equation (QME).

For any periodic driving system, we can derive a Floquet LvN equation which describes the EOM in the Floquet representation\cite{mosallanejad2023floquet}.
Here, we follow Ref. \citenum{mosallanejad2023floquet} to construct Floquet Hamiltonian ($\hat{H}^F$). In the following, we briefly introduce two main operators, which are the Fourier number operators $\hat{N}$ and the Fourier ladder operators $\hat{L}_n$. They have following properties,
\begin{equation}
    \hat{N}\ket{n}=n\ket{n}, \hat{L}_n\ket{m}=\ket{n+m}
\end{equation}
where $\ket{n}$ is the basis set in the Fourier space.
Then the Hamiltonian and density operator in Floquet representation would be
\begin{equation}
    \hat{H}^F = \sum_n\hat{H}^{(n)}\hat{L}_n + \hat{N}\hbar\omega
\end{equation}
\begin{equation}
    \hat{\rho}^F(t) = \sum_n\hat{\rho}^{(n)}(t)\hat{L}_n
\end{equation}
where $\hat{H}^{(n)}$ and $\hat{\rho}^{(n)}(t)$ are the Fourier expansion coefficients in $\hat{H}(t)=\sum_n\hat{H}^{(n)}e^{in\omega t}$ and $\hat{\rho}(t)=\sum_n\hat{\rho}^{(n)}(t)e^{in\omega t}$. By employing such definitions, the EOM of $\hat{\rho}^F(t)$ now reads as
\begin{equation}\label{F-EOM0}
    \frac{\partial}{\partial t}\hat{\rho}^F(t) = -\frac{i}{\hbar}[\hat{H}^F,\hat{\rho}^F(t)]
\end{equation}
Correspondingly, the Redfield equation for the molecule density operator in Floquet representation, $\hat{\rho}_s^F(t)$ reads as
\begin{equation}\label{F-EOM}
    \frac{\partial}{\partial t}\hat{\rho}_s^F(t) = -\frac{i}{\hbar}[\hat{H}_s^F,\hat{\rho}_s^F(t)] - \doublehat{\mathcal{L}}_{bs}^F\hat{\rho}_s^F(t)
\end{equation}
We refer to Eq. (\ref{F-EOM}) as the quantum master equation in Floquet representation (FR-QME).

We then proceed to perform a partial Wigner transformation for the density operator $\hat{\rho}_s^F(t)$, which is defined as 
\begin{equation}
    \hat{\rho}^F_{sW}(\boldsymbol{R,P},t) \equiv (2\pi\hbar)^{-N_{\alpha}}\int d\boldsymbol{X}\bra{\boldsymbol{R}-\boldsymbol{X}/2}\hat{\rho}^F_s(\boldsymbol{\hat{R},\hat{P}},t)\ket{\boldsymbol{R}+\boldsymbol{X}/2}e^{i\boldsymbol{P}\cdot\boldsymbol{R}/\hbar}
\end{equation}
where $\boldsymbol{R}$ and $\boldsymbol{P}$ can be interpreted as position and momentum variable in the classical limit instead of operators. $\boldsymbol{X}$ is the dummy variable.
$N_{\alpha}$ is the number of nuclear DoFs.
After performing a partial Wigner transformation for Eq. (\ref{F-EOM}),
we arrive at a Floquet representation based quantum-classical Liouville equation-classical master equation (FR-QCLE-CME),
\begin{equation}\label{FQME}
\begin{split}
    \frac{\partial}{\partial t}\hat{\rho}_{sW}^F(\boldsymbol{R,P},t) =& \frac{1}{2}\{\hat{H}_{sW}^F(\boldsymbol{R,P}),\hat{\rho}_{sW}^F\} - \frac{1}{2}\{\hat{\rho}_{sW}^F,\hat{H}_{sW}^F(\boldsymbol{R,P})\} \\&
    - \frac{i}{\hbar}[\hat{H}_{sW}^F,\hat{\rho}_{sW}^F] - \doublehat{\mathcal{L}}^F_{bsW}(\boldsymbol{R})\hat{\rho}^F_{sW}(t)
\end{split}
\end{equation}
here, $\{\cdot,\cdot\}$ is the Poisson bracket,
\begin{equation}
    \{A,B\} = \sum_{\alpha}\left(\frac{\partial A}{\partial R_{\alpha}}\frac{\partial B}{\partial P_{\alpha}} - \frac{\partial A}{\partial P_{\alpha}}\frac{\partial B}{\partial R_{\alpha}}\right)
\end{equation}
and $\hat{H}_{sW}^F$ indicates the partial Wigner transformation of $\hat{H}_s^F$,
The superoperator $\doublehat{\mathcal{L}}^F_{bsW}(\boldsymbol{R})$ in Eq. (\ref{FQME}) becomes
\begin{equation}
\begin{split}
    \doublehat{\mathcal{L}}^F_{bsW}(\boldsymbol{R})\hat{\rho}_{sW}^F(\boldsymbol{R,P},t) =& \frac{1}{\hbar^2}\int_0^{\infty}d\tau e^{-i\hat{H}_{sW}^Ft/\hbar}Tr_b([\hat{H}_{IcW}^F(t), [\hat{H}_{IcW}^F(t-\tau),\\&
    e^{i\hat{H}_{sW}^Ft/\hbar}\hat{\rho}_{sW}^F(t)e^{-i\hat{H}_{sW}^Ft/\hbar}\otimes\hat{\rho}_b^{eq}]])e^{i\hat{H}_{sW}^Ft/\hbar} 
\end{split}
\end{equation}

For surface hopping, it is useful to express FR-QCLE-CME in an adiabatic Floquet basis $\ket{\Psi_N^{F(ad)}}$, where $\hat{H}_{sW}^F\ket{\Psi_N^{F(ad)}}=\Tilde{E}_N\ket{\Psi_N^{F(ad)}}$. Then the FR-QCLE-CME in Eq. \ref{FQME} can be written as
\begin{equation}
\begin{split}
    \frac{\partial}{\partial t} \hat{\rho}_{sW}^{F(ad)}(\boldsymbol{R,P},t) =&
    -\frac{i}{\hbar} \Delta\hat{\Lambda}^F(\boldsymbol{R}) \circ \hat{\rho}_{sW}^{F(ad)}
    -\sum_{\alpha} \frac{P_{\alpha}}{m_{\alpha}}[\hat{D}_{\alpha}(\boldsymbol{R}), \hat{\rho}_{sW}^{F(ad)}] \\ &
    -\frac{1}{2}\sum_{\alpha} [\hat{F}_{\alpha}(\boldsymbol{R}), \frac{\partial \hat{\rho}_{sW}^{F(ad)}}{\partial P_{\alpha}}] - \sum_{\alpha} \frac{P_{\alpha}}{m_{\alpha}} \frac{\partial \hat{\rho}_{sW}^{F(ad)}}{\partial R_{\alpha}} \\ &
    - \doublehat{\mathcal{L}}^{F(ad)}_{bsW}(\boldsymbol{R})\hat{\rho}_{sW}^{F(ad)}
\end{split}
\end{equation}
where $\Delta\hat{\Lambda}^F$ is the eigenvalue difference matrix of $\hat{H}_{sW}^F$ ($\Delta\Lambda^F_{NM}=\Tilde{E}_N-\Tilde{E}_M$), $A\circ B$ is Hadamard product of two matrices with same dimension,  $\hat{F}^{\alpha}$ is the force matrix ($F_{NM}^{\alpha}\equiv -\bra{\Psi_N^{F(ad)}}\frac{\partial\hat{H}_{sW}^F}{\partial R^{\alpha}}\ket{\Psi_M^{F(ad)}}$), and $\hat{D}^{\alpha}$ is the derivative coupling matrix ($D^{\alpha}_{NM}\equiv F^{\alpha}_{NM}/(\Tilde{E}_N-\Tilde{E}_M)$).
In Appendix A, we give a form of  the Redfield operator $\doublehat{\mathcal{L}}^F_{bsW}(\boldsymbol{R})$ both in diabatic and adiabatic representations.
Next, we will use trajectory-based algorithms to solve the FR-QCLE-CME.


\subsection{B. Floquet representation based Surface hopping (FR-SH) algorithm}
The FR-QCLE-CME can be solved by a Floquet representation based surface hopping (FR-SH) algorithm.
Similar to the SH proposed by Dou et al.\cite{dou2017generalized}, for each trajectory, we propagate the density matrix $\hat{\sigma}$ according to 
\begin{equation}\label{rho_dot}
\begin{split}
   \dot{\hat{\sigma}}^{F(ad)}=&
   -\frac{i}{\hbar}\Delta\hat{\Lambda}^F(\boldsymbol{R})\circ \hat{\sigma}^{F(ad)}
    -\sum_{\alpha}\frac{P_{\alpha}}{m_{\alpha}}[\hat{D}_{\alpha}(\boldsymbol{R}), \hat{\sigma}^{F(ad)}] - \doublehat{\mathcal{L}}^{F(ad)}_{bsW}(\boldsymbol{R})\hat{\sigma}^{F(ad)}
\end{split}
\end{equation}
as well as position and momentum ($\boldsymbol{R}$ and $\boldsymbol{P}$) on the active potential surface $\lambda$,
\begin{equation}\label{Rdot}
    \dot{R}_{\alpha} = \frac{P_{\alpha}}{m_{\alpha}}
\end{equation}
\begin{equation}\label{Pdot}
    \dot{P}_{\alpha} = F_{\alpha({\lambda\lambda})}
\end{equation}
In the spirit of Tully's surface hopping, the nuclei hops among adiabatic potential energy surfaces. The total change of the population on state $M$ is 
\begin{equation}\label{tully}
\begin{split}
    \dot{\sigma}_{MM}^{F(ad)}=&
    -\sum_{\alpha K}\frac{P_{\alpha}}{m_{\alpha}}(D_{{\alpha}(MK)}(\boldsymbol{R})\sigma_{KM}^{F(ad)}-\sigma_{MK}^{F(ad)}D_{{\alpha}(KM)}(\boldsymbol{R})) - \sum_{KL}\mathcal{L}_{MM,KL(bsW)}^{F(ad)}(\boldsymbol{R})\sigma_{KL}^{F(ad)}
\end{split}
\end{equation}
The first term on the right hand side (RHS) of Eq. (\ref{tully}) indicates hopping due to derivative coupling $D_{NM}^{\alpha}$. Therefore, we can define the hopping rate $k_{N\rightarrow M}^D$ caused by derivative coupling as
\begin{align}\label{kd}
    k_{N\rightarrow M}^D = \Theta\left(-2Re\sum_{\alpha}\frac{P_{\alpha}}{m_{\alpha}}\frac{D_{{\alpha}(MN)}\sigma_{NM}^{F(ad)}}{\sigma_{NN}^{F(ad)}}\right)
\end{align}
where $\Theta$ function is defined as
\begin{align}
    \Theta(x) =
    \begin{cases}
        x, (x\geq0) \\
        0, (x<0)
    \end{cases}
\end{align}
The second term on the RHS of Eq. (\ref{tully}) is an extra hopping due to molecule-metal interaction. This term has both diagonal and off-diagonal contributions:
\begin{equation}
\begin{split}
    -\sum_{KL}\mathcal{L}_{MM,KL(bsW)}^{F(ad)}(\boldsymbol{R})\sigma_{KL}^{ad} =& -\sum_{N}\mathcal{L}_{MM,NN(bsW)}^{F(ad)}(\boldsymbol{R})\sigma_{NN}^{ad}
    - \sum_{K\neq L}\mathcal{L}_{MM,KL(bsW)}^{F(ad)}(\boldsymbol{R})\sigma_{KL}^{ad}
\end{split}
\end{equation}
Here, we employ the secular approximation to ignore the off-diagonal part, 
which is accurate in long time dynamics.
Such that the hopping rate caused by interactions between the system and electronic bath $k_{N\rightarrow M}^{\mathcal{L}}$ reads as
\begin{equation}\label{kL}
\begin{split}
    k_{N\rightarrow M}^{\mathcal{L}} = -\mathcal{L}_{MM,NN(bsW)}^{F(ad)}
\end{split}
\end{equation}
The details of operating such FR-SH algorithm step by step are given in Appendix B.

\section{3. Results}
To test our FR-SH algorithm, we use a donor-acceptor-metal model. We consider here the condition of external periodic drivings acting on the coupling strength between donor and acceptor, namely the case of light interacting with transition dipole moment of donor and acceptor. The system Hamiltonian corresponding to this case is
\begin{equation}\label{H1}
\begin{split}
    \hat{H}_{sW}(x,t) = & E_D(x)\hat{d}_D^+\hat{d}_D+E_A(x)\hat{d}_A^+\hat{d}_A + (W+A\sin{(\Omega t)})(\hat{d}_D^+\hat{d}_A + \hat{d}_A^+\hat{d}_D) \\ &
    + \frac{1}{2}m\omega^2x^2 + \frac{p^2}{2m}
\end{split}
\end{equation}
where $A$ is the driving amplitude and $\Omega$
is the driving frequency.
We further set $E_D(x)=gx\sqrt{2m\omega/\hbar}+\epsilon_D$ and $E_A(x)=0$.
In this model, we have
\begin{equation}
\begin{split}
    \Gamma_{AA} = \Gamma, \Gamma_{DD}=\Gamma_{DA}=\Gamma_{AD} = 0
\end{split}
\end{equation}

We benchmark the FR-SH algorithm against FR-QME in Eq. \ref{F-EOM}.
Note that we are dealing with a four-state system in Fock space: diabatic state 1 is both donor and acceptor are unoccupied; diabatic state 2(3) is only the donor(acceptor) is occupied; and diabatic state 4 is both donor and acceptor are occupied. 
In Floquet representation, all these four states in Fock space will be expanded in Floquet space. Note that we need to truncate the Floquet space according to the ratio of driving amplitude $A$ and driving frequency $\Omega$, see details in Ref. \citenum{wang2023nonadiabatic}.
Briefly, the greater the $\frac{A}{\Omega}$ is, the larger the Floquet space should be.
We prepare the nuclei with a Boltzmann distribution on diabatic donor states (diabatic state 2). 
To transfer from adiabatic to diabatic states, we follow the scheme in Refs. \citenum{landry2013communication} and \citenum{dou2017generalized}, such that the diabatic donor population $\expval{\hat{d}^+_D\hat{d}_D}$ is given,
\begin{equation}
\begin{split}
    \expval{\hat{d}^+_D\hat{d}_D}=\frac{1}{N_{traj}}\sum_{l}^{N_{traj}}\left(\sum_{i}\abs{U_{ai}}^2\delta_{i,\lambda^l} + \sum_{i<j}2\Re(U_{ai}\sigma_{ij}^{F(l)}U_{aj}^{*}) + \sum_i\abs{U_{bi}}^2\delta_{i,\lambda^l} \right)
\end{split}
\end{equation}
here, $N_{traj}$ is the total number of trajectories, which is $10,000$ in this study, 
$a$ is an index for diabatic state 2 which only donor is occupied, $b$ is an index for diabatic state 4 which both donor and acceptor are occupied, $l$ is an index for trajectories, and $\lambda$ is the active Floquet potential energy surface.  
For FR-QME, we use $100$ phonons to get good convergences.

In the following, we perform the dynamics of diabatic donor populations and nuclear kinetics energy. 
We choose three kinds of driving amplitudes: weak driving ($A=0.005$), medium strong driving ($A=0.01$), and strong driving ($A=0.02$), comparing with nuclear oscillation strength ($\hbar\omega=0.003$). All these drivings are fast drivings that we fix a relatively large driving frequency $\Omega=0.1$.
We benchmark the FR-SH algorithm with FR-QME.

\begin{figure}
    \centering
    \includegraphics[scale=0.08]{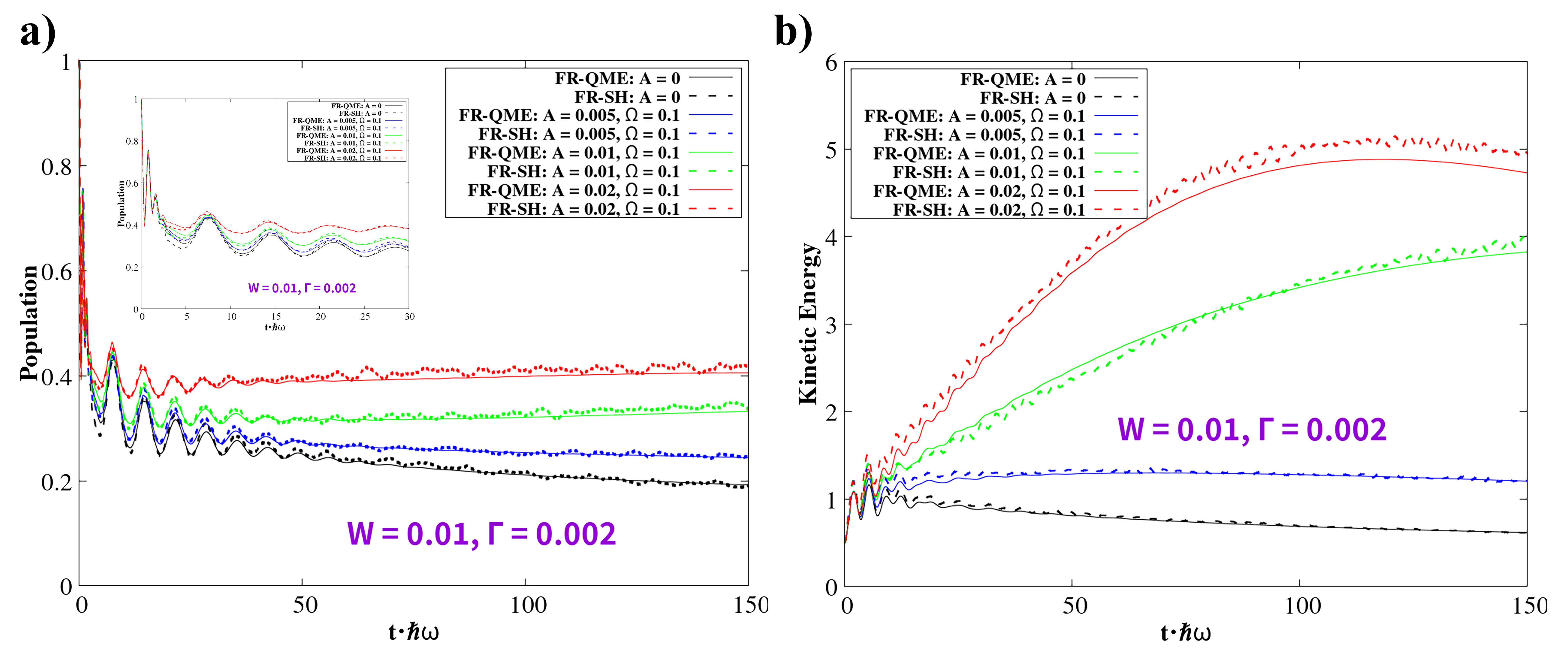}
    \caption{Diabatic electronic population on the donor $\expval{\hat{d}_D^+\hat{d}_D}$ and kinetic energy as a function of time under Floquet drivings on donor-acceptor coupling. $kT = 0.01$, $\hbar\omega = 0.003$, $g = 0.0075$, $\epsilon_D = 2E_r$, $E_r = g^2/(\hbar\omega)$, $W = 0.01$, and $\Gamma = 0.002$.}
    \label{fig:strong}
\end{figure}

In Fig. \ref{fig:strong}, we work in the regime where $W$ is much larger than $\Gamma$.
In such a case, the FR-SH (dash line) agrees well with FR-QME (solid line) under any drivings, for both diabatic population (Fig. \ref{fig:strong}a) and kinetic energy (Fig. \ref{fig:strong}b) dynamics.
It is because $k_{N\rightarrow M}^{\mathcal{L}}$ (relate to $\Gamma$) is much smaller than $k_{N\rightarrow M}^D$ (relate to $W$). 
That is to say, the ignorance of off-diagonal terms in hopping rates $k_{N\rightarrow M}^{\mathcal{L}}$ cannot affect dynamics in both short and long times.
We can see from long time dynamics in Fig. \ref{fig:strong}a that with increasing the driving amplitude $A$, the electronic population of donor reaches to a higher steady state.
Additionally, with increasing the driving amplitude, the system reaches a higher temperature as shown in Fig. \ref{fig:strong}b, which is the heating effect arising from the external drivings\cite{wang2023nonadiabatic,wang2023EF}.

\begin{figure}
    \centering
    \includegraphics[scale=0.08]{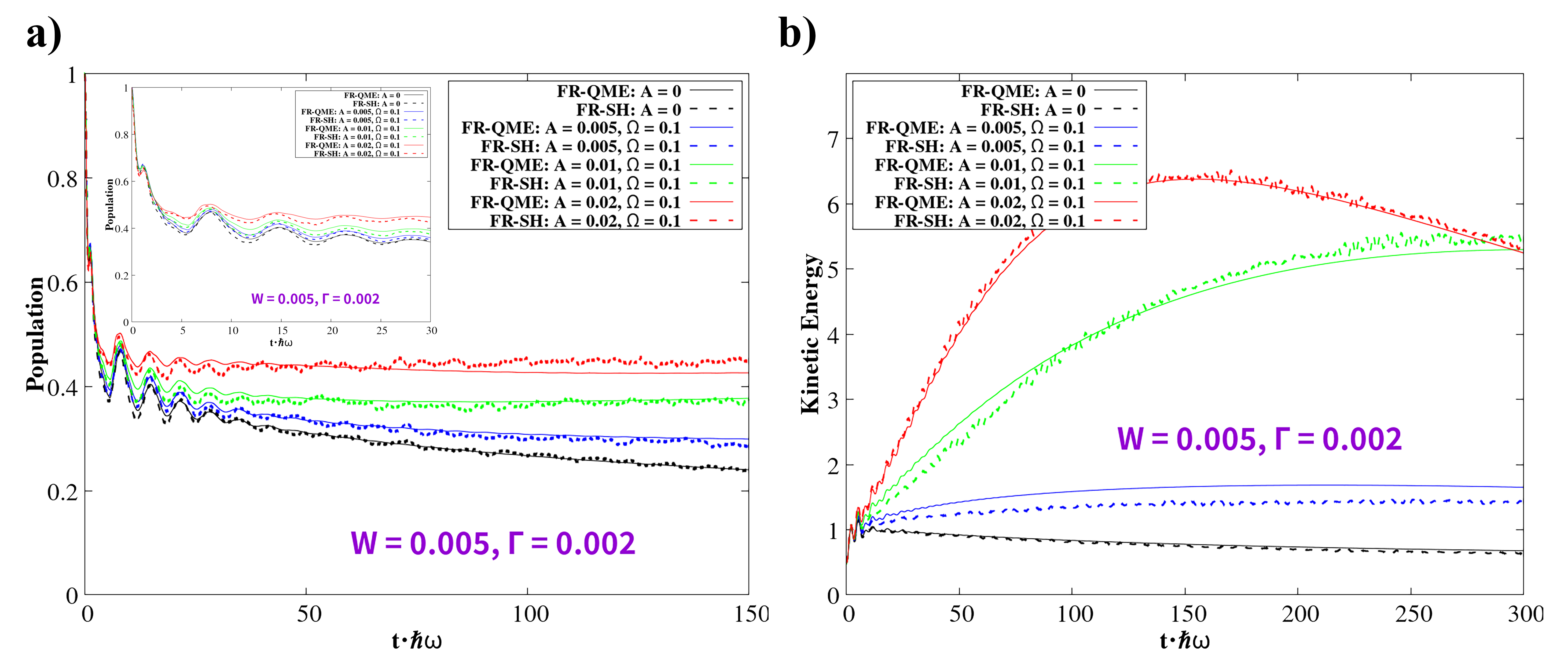}
    \caption{Diabatic electronic population on the donor $\expval{\hat{d}_D^+\hat{d}_D}$ and kinetic energy as a function of time under Floquet drivings on donor-acceptor coupling. $kT = 0.01$, $\hbar\omega = 0.003$, $g = 0.0075$, $\epsilon_D = 2E_r$, $E_r = g^2/(\hbar\omega)$, $W = 0.005$, and $\Gamma = 0.002$.}
    \label{fig:moderate}
\end{figure}

Next, we work in the regime where $W$ is little larger than $\Gamma$ (Fig. \ref{fig:moderate}). In such a case, FR-SH covers right steady state in the long time dynamics as compared with FR-QME. However, FR-SH fails at the short time dynamics (as shown in inset of Fig. \ref{fig:moderate}a).
It is because the off-diagonal terms in $k_{N\rightarrow M}^{\mathcal{L}}$ become more important in this case, such that the ignorance of them can affect initial dynamics.
Again, the electronic population of donor reaches to a higher steady state with increasing the driving amplitude $A$ from Fig. \ref{fig:moderate}a.
And the system reaches to higher temperature as $A$ increased (see Fig. \ref{fig:moderate}b).

\begin{figure}
    \centering
    \includegraphics[scale=0.08]{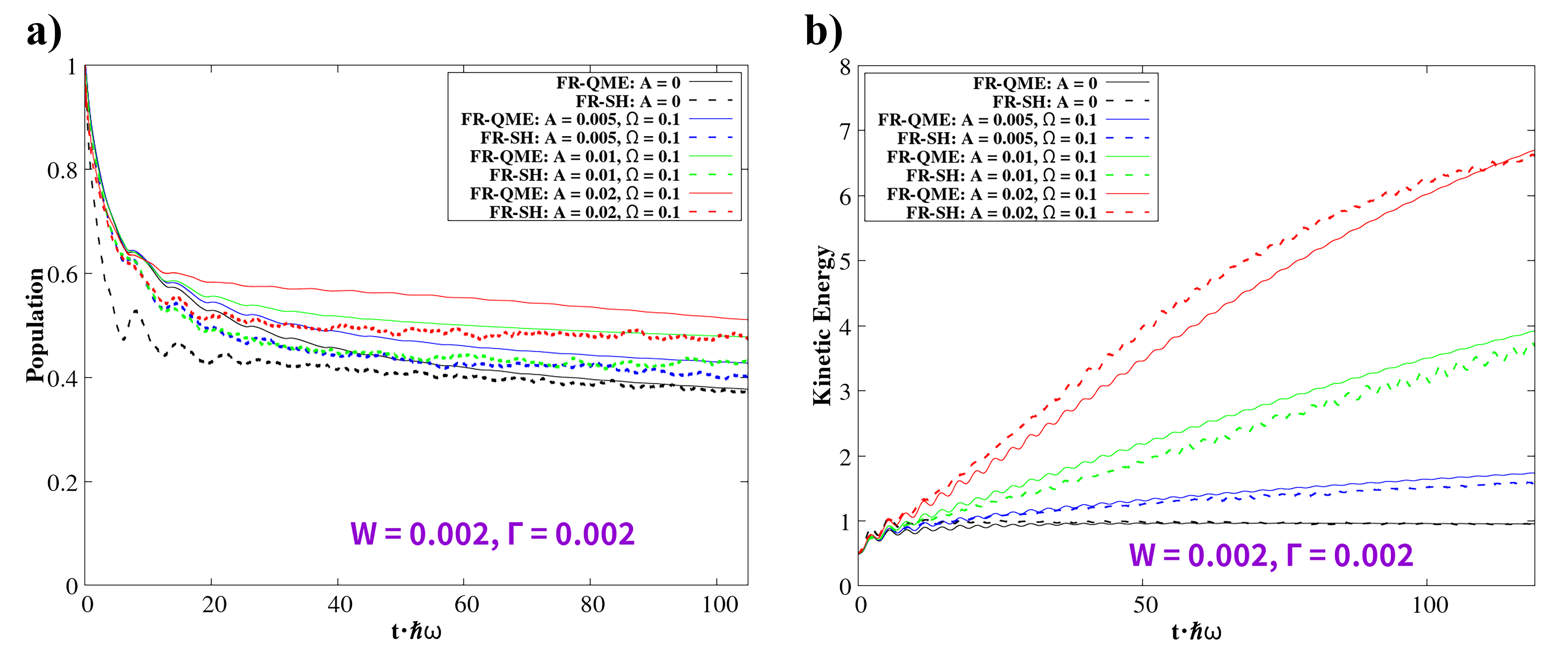}
    \caption{Diabatic electronic population on the donor $\expval{\hat{d}_D^+\hat{d}_D}$
    and kinetic energy as a function of time under Floquet drivings on donor-acceptor coupling. $kT = 0.01$, $\hbar\omega = 0.003$, $g = 0.0075$, $\epsilon_D = 2E_r$, $E_r = g^2/(\hbar\omega)$, $W = 0.002$, and $\Gamma = 0.002$.}
    \label{fig:weak}
\end{figure}

Finally, when $W$ is comparable with $\Gamma$ (Fig. \ref{fig:weak}), we see differences between FR-SH and FR-QME under any kinds of drivings. In such a case, the off-diagonal terms in $k_{N\rightarrow M}^{\mathcal{L}}$ are dominant. Therefore, secular approximated surface hopping algorithm here cannot cover the right dynamics \cite{dou2017generalized}.

Overall, our generalized RH-SH method works well under fast drivings when $W$ is much larger than $\Gamma$.

\section{4. Conclusion}

In summary, we have proposed a generalized Floquet representation based surface hopping (FR-SH) algorithm to deal with the nonadiabatic dynamics near metal surface with fast periodic drivings.
In the regime of donor-acceptor coupling $W$ is relatively larger than system-bath coupling $\Gamma$, FR-SH agrees well with FR-QME for both diabatic population and nuclear kinetic energy dynamics under different Floquet drivings.
However, when $W$ is comparable with $\Gamma$, FR-SH fails due to its ignorance of the off-diagonal terms in $k_{N\rightarrow M}^{\mathcal{L}}$.
We see that with increasing the driving amplitude $A$, the electronic population of donor reaches to a higher steady state, and the nuclear kinetic energy reaches to a higher temperature.
We expect this generalized FR-SH algorithm to be useful for modeling realistic nonadiabtic dynamics near metal surface with periodic drivings.

\section{Appendix A. The Redfield operator}\label{redfield}
A simplified form for the Redfield operator $\doublehat{\mathcal{L}}^F_{bsW}(\boldsymbol{R})$ can be expressed in either diabatic or adiabatic representation.
In the diabatic representation, it is
\begin{equation}
\begin{split}
    \doublehat{\mathcal{L}}^F_{bsW}\hat{\rho}_{sW}^F =& 
    \sum_{nm}\frac{\Gamma_{mn}}{2\hbar}\hat{d}_m^{F+}\hat{U}\Tilde{\mathbb{D}}_n^F\hat{U}^+\hat{\rho}_{sW}^F(t) \\&
    + \sum_{nm}\frac{\Gamma_{mn}}{2\hbar}\hat{d}_m^F\hat{U}\mathbb{D}_n^{F+}\hat{U}^+\hat{\rho}_{sW}^F(t) \\&
    - \sum_{nm}\frac{\Gamma_{mn}}{2\hbar}\hat{d}_m^{F+}\hat{\rho}_{sW}^F(t)\hat{U}\mathbb{D}_n^F\hat{U}^+ \\&
    - \sum_{nm}\frac{\Gamma_{mn}}{2\hbar}\hat{d}_m^F\hat{\rho}_{sW}^F(t)\hat{U}\Tilde{\mathbb{D}}_n^{F+}\hat{U}^+ + h.c.
\end{split}
\end{equation}
In the adiabatic representation, it is ($\hat{\rho}_{sW}^{F(ad)}=\hat{U}^+\hat{\rho}_{sW}^F\hat{U}$)
\begin{equation}
\begin{split}
    \doublehat{\mathcal{L}}^F_{bsW}\hat{\rho}_{sW}^{F(ad)} =& 
    \sum_{nm}\frac{\Gamma_{mn}}{2\hbar}\hat{U}^+\hat{d}_m^{F+}\hat{U}\Tilde{\mathbb{D}}_n^F\hat{\rho}_{sW}^{F(ad)}(t) \\&
    + \sum_{nm}\frac{\Gamma_{mn}}{2\hbar}\hat{U}^+\hat{d}_m^F\hat{U}\mathbb{D}_n^{F+}\hat{\rho}_{sW}^{F(ad)}(t) \\&
    - \sum_{nm}\frac{\Gamma_{mn}}{2\hbar}\hat{U}^+\hat{d}_m^{F+}\hat{U}\hat{\rho}_{sW}^{F(ad)}(t)\mathbb{D}_n^F \\&
    - \sum_{nm}\frac{\Gamma_{mn}}{2\hbar}\hat{U}^+\hat{d}_m^F\hat{U}\hat{\rho}_{sW}^{F(ad)}(t)\Tilde{\mathbb{D}}_n^{F+} + h.c.
\end{split}
\end{equation}
here, $(\mathbb{D}_n)_{NM}\equiv(\hat{U}^+\hat{d}_n\hat{U})_{NM}f(\Tilde{E}_N-\Tilde{E}_M)$, $(\mathbb{D}_n^+)_{NM}\equiv(\hat{U}^+\hat{d}_n^+\hat{U})_{NM}f(\Tilde{E}_N-\Tilde{E}_M)$, $(\Tilde{\mathbb{D}}_n)_{NM}\equiv(\hat{U}^+\hat{d}_n\hat{U})_{NM}(1-f(\Tilde{E}_N-\Tilde{E}_M))$, $(\Tilde{\mathbb{D}}_n^+)_{NM}\equiv(\hat{U}^+\hat{d}_n^+\hat{U})_{NM}(1-f(\Tilde{E}_N-\Tilde{E}_M))$, where $\hat{U}$ and $\Tilde{E}_N$ are the eigenvectors and eigenvalues of the Floquet system Hamiltonian $\hat{H}^F_{s(el)}$, respectively. $f(E)=1/(e^{E/(kT)}+1)$ is the Fermi function.

\section{Appendix B. FR-SH algorithm step by step}\label{FSH-step}
The procedure of FR-SH is largely similar with that in Ref. \citenum{dou2017generalized}. The only difference is that we are now in Floqeut states. 
The FR-SH algorithm is performed as follows:
\begin{enumerate}
    \item Prepare the initial $\hat{\sigma}$, $\boldsymbol{R}$, and $\boldsymbol{P}$. Choose an active Floquet PES (say $\lambda=N$).
    \item Evolve $\hat{\sigma}$, $\boldsymbol{R}$, and $\boldsymbol{P}$ according to Eqs. (\ref{rho_dot}-\ref{Pdot}) for a time step $\Delta t$ on the active Floquet PES ($\lambda=N$).
    \item Calculate the hopping rate $k_{N\rightarrow M}^D$ and $k_{N\rightarrow M}^{\mathcal{L}}$ for all Floquet PESs based on Eqs. (\ref{kd}) and (\ref{kL}). Generate a random number $\xi\in[0,1]$. Then we define a total rate as $S_N^M=\sum_{K=1}^Mk_{N\rightarrow K}^{total}$. The total number of PESs is $T$
    \begin{itemize}
        \item If $\xi>S_N^T$, the nuclei remain on surface $N$.
        \item Else if $S_N^{M-1}\Delta t<\xi<(S_N^{M-1}+k_{N\rightarrow M}^{\mathcal{L}})\Delta t$, the nuclei hops to surface $M$ ($\lambda=M$), without momentum rescaling.
        \item Else if $(S_N^{M-1}+k_{N\rightarrow M}^{\mathcal{L}})\Delta t<\xi<S_N^M\Delta t$, the nuclei hops to PES $M$ ($\lambda=M$) with momentum rescaling along the direction of the derivative coupling:
        \begin{equation}
            \boldsymbol{P}_{new} = \boldsymbol{P} + \kappa\boldsymbol{D}_{NM}|\boldsymbol{D}_{NM}|
        \end{equation}
        and energy conservation implies
        \begin{equation}
        \begin{split}
            & \sum_{\alpha}\frac{(P_{\alpha,new})^2}{2m_{\alpha}} + \Tilde{E}_N(\boldsymbol{R}) \\&
            = \sum_{\alpha}\frac{(P_{\alpha})^2}{2m_{\alpha}} + \Tilde{E}_M(\boldsymbol{R})
        \end{split}
        \end{equation}
        here, we only choose the root with smaller $|\kappa|$.
    \end{itemize}
    \item Repeat step 2 and 3 for the desired number of time steps.
\end{enumerate}

\begin{acknowledgement}

This material is based upon work supported by National Science Foundation of China (NSFC No. 22273075)

\end{acknowledgement}


\bibliography{achemso-demo}

\end{document}